\documentclass[12pt,a4paper]{article}

\usepackage{amsfonts,amssymb,amsmath,amsopn,amsthm,graphicx}

 \numberwithin{equation}{section}

\newtheorem{thm}{Theorem}[section]
\newtheorem{defin}{Definition}[section]
\newtheorem{lem}{Lemma}[section]

\newtheorem{prop}{Proposition}[section]
\newtheorem{rem}{Remark}[section]

\newcommand{\be}{\begin{equation}}
\newcommand{\ee}{\end{equation}}
\newcommand{\bea}{\begin{eqnarray}}
\newcommand{\eea}{\end{eqnarray}}
\newcommand{\eps}{\epsilon}

\newcommand{\pd}{\partial}
\newcommand{\D}{\,\textrm{d}}

\newcommand{\N}{\mathbb{N}}

\newcommand{\R}{\mathbb{R}}
\newcommand{\C}{\mathbb{C}}

\newcommand{\ra}{\rightarrow}

\begin{document}

\title{On the Exponential Decay of Magnetic Stark Resonances}
\author{Christian Ferrari$^a$ and Hynek Kova\v r\'{\i}k$^{b}$}
\date{}
\maketitle

\begin{flushleft}
 {\em a) Institute for Theoretical Physics, Ecole
 Polytechnique F\'ed\'erale de Lausanne, CH-1015  Lausanne, Switzerland
 \\ b) Institut f\"ur Analysis, Dynamik und Modellierung, Universit\"at
 Stuttgart, Pfaffenwaldring 57, D-70569 Stuttgart, Germany}\footnote[1]{also
    on leave from  Department of Theoretical Physics, Nuclear Physics
    Institute, Academy of Sciences, 25068 \v{R}e\v{z} near Prague, Czech Republic}
\end{flushleft}

\begin{abstract}
We study the time decay of magnetic Stark resonant states. As our
main result we prove that for sufficiently large time these states
decay exponentially with the rate given by the imaginary parts of
eigenvalues of certain non-selfadjoint operator. The proof is
based on the method of complex translations.
\end{abstract}

\section{Introduction}

\noindent The purpose of this paper is to study the decay
properties of resonances in two dimensions in the presence of
crossed magnetic and electric fields and a potential type
perturbation. We assume that the magnetic field acts in the
direction perpendicular to the electron plane with a constant
intensity $B$ and that the electric field of constant intensity
$F$ points in the $x-$direction. The perturbation $V(x,y)$ is
supposed to satisfy certain localisation conditions. The
corresponding quantum Hamiltonian reads as follows
$$
 H(F) = H(0)-Fx=H_L+V-Fx,
$$
where $H_L$ is the Landau Hamiltonian of an electron in a
homogeneous magnetic field of intensity $B$.
\par
We begin with the definition of a resonance in terms of an
exponential time decay of the corresponding resonant states. In
Section \ref{expdecay} we show the connection between these time
decaying states and the usual spectral deformation notion of
resonance. The basic mathematical tool we use is the method of
complex translations for Stark Hamiltonians, which was introduced
in \cite{AH} as a modification of the original theory of complex
scaling \cite{AC}, \cite{BC}. Following \cite{AH} we consider the
transformation $U(\theta)$, which acts as a translation in
$x-$direction; $(U(\theta)\psi)(x)=\psi(x+\theta)$. For non real
$\theta$ the translated operator $H(F,\theta)=U(\theta)H(F)
U^{-1}(\theta)$ is non-selfadjoint and therefore can have some
complex eigenvalues. The main result of Section \ref{expdecay},
Theorem \ref{lifetime}, tells us that if $\phi$ is an
eigenfunction of $H(0)$, then $(\phi,e^{-itH(F)}\, \phi)$ decays
exponentially at the rate given by the imaginary parts of the
eigenvalues of $H(F,\theta)$. Theorem \ref{lifetime} thus can be
regarded as a generalisation of the result obtained in
\cite{Herbst}, where the exponential decay was proved for the
Stark Hamiltonians without magnetic field.
\par
Of course on would like to know how the resonance widths behave as
functions of $F$. This question is discussed in the
forthcoming paper, in which we prove that for $F\to 0$ the resonance widths
decay as $\exp[-\frac{B}{F^2}]$ in contrast with the usual Stark resonances, 
where the behaviour is exponential. However, the technique
used in our next paper requires some specific properties of the Green
function $G_1({\bf  x},{\bf x}';z)$ of the operator
$$
H_1(F)= H_L-Fx,
$$
in the limit $F\to 0$. In particular, on need to know that
$G_1({\bf  x},{\bf x}';z)$
is exponentially decaying with respect to $(x'-x)^2$ and $|y'-y|$.
While similar behaviour is well known in case of purely magnetic
Hamiltonian, where the Green function is given explicitly, to the
best of our knowledge there is no explicit formula for the Green
function of the crossed fields Hamiltonian $H_1(F)$. The direct application of
these results on the crossed fields Green function motivates us to include them
as a second part of the present paper. However, the estimations of
$G_1({\bf  x},{\bf x}';z)$ could be of general interest for other problems
dealing with simultaneous electric and magnetic fields.

\section{The Model}

We work in the system of units, where $m=1/2,\, e=1,\, \hbar =1$.
The crossed fields Hamiltonian is then given by \be H_1(F)= H_L-Fx
= (-i\pd_x+By)^2 -\pd_y^2-Fx,\quad {\rm on}\quad L^2(\R^2). \ee
Here we use the Landau gauge with ${\bf A}(x,y)=(-By,0)$. A
straightforward application of \cite[Thm. X.37]{rees} shows that
$H_1(F)$ is essentially self-adjoint on $C_0^{\infty}(\R^2)$, see
also \cite[Prob. X.38]{rees}. Moreover, one can easily check that
\begin{equation}
\sigma(H_1(F))=\sigma_{ac}(H_1(F))=\R
\end{equation}
As mentioned in the Introduction we employ the translational
analytic method developed in \cite{AH}. We introduce the
translated operator $H_1(F,\theta)$ as follows:
\begin{equation}
H_1(F,\theta)= U(\theta)H_1(F)U^{-1}(\theta)
\end{equation}
where
\begin{equation}
\left(U(\theta)f\right)(x,y):=\left(e^{ip_x\theta}f\right)(x,y)=f(x+\theta,
y)
\end{equation}
An elementary calculation shows that
\begin{equation}
H_1(F,\theta)=H_1(F)-F\theta
\end{equation}
Operator $H_1(F,\theta)$ is clearly analytic in $\theta$.
Following \cite{AH} we define the class of $H_1(F)-$translation
analytic potentials.

\begin{defin}
Suppose that $V(z,y)$ is analytic in the strip $|\Im z|<\beta$,
$\beta>0$ independent of $y$. We then say that $V$ is
$H_1(F)-$translation analytic if $V(x+z,y)(H_1(F)+i)^{-1}$ is a
compact analytic operator valued function of $z$ in the given
strip.
\end{defin}

We can thus formulate the conditions to be imposed on $V$:
\begin{itemize}
\item
[$(a)$] $V(x,y)$ is $H_1(F)-$translation analytic in the strip
$|\Im z|<\beta$.
\item
[$(b)$] There exists $\beta_0\leq \beta$ such that for $|\Im
z|\leq \beta_0$ the function $V(x+z,y)$ is uniformly bounded and
$$\lim_{x,y\ra\pm\infty}|V(x+z,y)|=0$$
\item[(\emph{c})]
The operator $H(F)=H_1(F)+V$ has purely absolutely continuous
spectrum.
\end{itemize}

In order to characterise the potential class for which the above
conditions are fulfilled let us assume for the moment, that the
integral kernel of $(H_1(F)+i)^{-1}$ has a local logarithmic
singularity at the origin. This is a very plausible hypothesis,
see Lemma \ref{shortdist}, it then follows that any $L^2(\R^2)$
function which tends to zero at infinity and can be analytically
continued in a given strip $|\Im z|<\beta$ satisfies the
conditions $(a)$ and $(b)$. We can take a Gaussian as an
elementary example. The condition $(c)$ is more delicate. Although
the quantum tunnelling phenomenon leads us to believe that all the
impurity states becomes unstable once the electric field is added,
there is no rigorous results on the potential class that satisfies
$(c)$.

From the well known perturbation argument, \cite{Kato}, we see
that under assumption $(b)$
\begin{equation}
H(F,\theta) = U(\theta)H(F)\, U^{-1}(\theta)= H_1(F,\theta)+V\left
(x+\theta,y\right )
\end{equation}
forms an analytic family of type $A$.\\  Furthermore, since
$V(x+\theta,y)(H_1(F)+i)^{-1}$ is compact by $(a)$, we have
\begin{equation}\label{specH1}
\sigma_{ess}(H(F,\theta)) = \sigma_{ess}(H_1(F,\theta)) = \R -ibF
\end{equation}
where $\theta=i b,\, b\in\R$. By standard arguments \cite[Prob. XIII.76]{rees}, all
eigenvalues of $H(F,ib)$ lie in the strip $-bF<\Im z\leq 0$ and
are independent of $b$ as long as they are not covered by the
essential spectrum.

\section{Exponential decay} \label{expdecay}

\noindent The resonant states for our model are defined in the
following way:

\begin{defin}
We say that $\varphi$ is a resonant state of $H(F)$ with width
$\Gamma$, if there exists some $\epsilon>0$, such that
$$
|(\varphi,e^{-itH(F)}\, \varphi)|^2 = e^{-t\, \Gamma}(1+R(t)),
$$
where
$$
|R(t)|=\mathcal{O}(e^{-t\, \epsilon}),\quad \text{as}\quad
t\ra\infty.
$$
\end{defin}

\noindent We remark that for a bounded below Hamiltonian the decay
law can be exponential only for times neither too small nor too
large, \cite{exner}. However, in our case, due to the fact that
$H(F)$ is unbounded from below, the above definition makes sense.
For a detailed discussion of the problem of definition of
resonance see also \cite{sim2}. The goal of this section is to
prove that the resonance width $\Gamma$ is given by an imaginary
part of the associated complex eigenvalue of $H(F,\theta)$. We
will borrow the ideas from \cite{Herbst} where a similar problem
in three dimensions was treated in the absence of magnetic field.
The main ingredient of our analysis is the proof of the fact that
$H(F,\theta)$ can have only a finite number of eigenvalues in a
given strip. We will need the following claim.

\begin{prop} \label{limit}
Let $f,g$ be bounded functions with compact support in $\R^2$.
Then
$$
\lim_{\lambda\ra\pm\infty}\|f(H_1(F)-\lambda-i\, \gamma)^{-1}g\|=0
$$
for $F\geq 0$ and uniformly for $\gamma$ in the compacts of
$\R\setminus \{0\}$.
\end{prop}

\begin{proof}
We take $\gamma<0$ and write\footnote{here $\N_0:=\N\cup\{0\}$ }
\begin{eqnarray} \label{resolventa}
f(H_1(F)-\lambda-i\, \gamma)^{-1}g &= & -i\int_0^{\infty}
(fe^{itH_1(F)}g)e^{\gamma t}e^{-i\lambda t}\D t
:= \int_0^{\infty} G(t)e^{-i\lambda t} \D t \nonumber\\
&= & \int_0^{\eps}G(t)e^{-i\lambda t}\D t+ \sum_{n\in\N}
\int_{n\pi/B- \eps}^{n\pi/B+\eps}
G(t)e^{-i\lambda t} \D t \nonumber\\
&+ &
\sum_{n\in\N_0}\int_{n\pi/B+\eps}^{(n+1)\pi/B-\eps}G(t)e^{-i\lambda
t} \D t
\end{eqnarray}
\noindent The first term on the right hand side is bounded from
above by $\|f\|_{\infty}\, \|g\|_{\infty}\, \eps$.
 For the second we have
$$
\left \|\sum_{n\in\N}
\int_{n\pi/B-\eps}^{n\pi/B+\eps}G(t)e^{-i\lambda t} \D t \right \|
\leq 2\eps\, \|f\|_{\infty}\,
\|g\|_{\infty}\sum_{n\in\N}e^{\gamma(n\pi/B-\eps)}
$$
which implies
\begin{eqnarray} \label{norm}
\|f(H_1(F)-\lambda-i\, \gamma)^{-1}g\| &\leq & \eps\,
\|f\|_{\infty}\,
\|g\|_{\infty}\left(\frac{2e^{-\gamma\epsilon}}{1-e^{\gamma\pi/B}}
+1\right ) \nonumber\\
&+ & \left \|
\sum_{n\in\N_0}\int_{n\pi/B+\eps}^{(n+1)\pi/B-\eps}G(t)e^{-i\lambda
t} \D t \right \|
\end{eqnarray}
All terms in the sum on the r.h.s. of (\ref{norm}) can be
integrated by parts to give
\begin{eqnarray}
\int_{n\pi/B+\eps}^{(n+1)\pi/B-\eps}G(t)e^{-i\lambda t} \D t & =
&\frac{1}{i\, \lambda}
\int_{n\pi/B+\eps}^{(n+1)\pi/B-\eps}G'(t)e^{-i\lambda t} \D t
\nonumber
\\
& - & \left [ \frac{1}{i\, \lambda}\, G(t)e^{-i\, \lambda t}\right
]_{n\pi/B+\eps}^{(n+1)\pi/B-\eps}
\end{eqnarray}
where the second term on the r.h.s. is bounded above by
$2\|f\|_{\infty}\, \|g\|_{\infty}|\lambda|^{-1}$. In order to
estimate the first term we use the integral kernel of the
evolution operator $e^{-i\, tH_1(F)}$ in the gauge where
$H_L=p_x^2+(p_y-Bx)^2$ (keeping in mind that the norm is
gauge-invariant). From the formula (\ref{evol}) given in Appendix
A we then deduce the integral kernel of $G'(t)$

\begin{eqnarray}
& &(x,y|G'(t)|x_0,y_0)=\frac{1}{2\pi i}\sqrt{\frac{B}{2}}e^{\gamma
t}f(x,y)g(x_0,y_0)e^{iS_{-t}[w_{cl}(\cdot)]}\frac{1}{\sin(Bt)}
\times
\nonumber\\
&\times& \Bigg\{\gamma + B\cot(Bt) +  \frac{i}{4}\Big(u^2 -
2F(x+x_0) -
\frac{B^2}{\sin^2(Bt)}[(x-x_0)^2+(y-y_0+ut)^2] \nonumber\\
&+& 2F\cot(Bt) (y-y_0+ut)\Big) \Bigg\}
\end{eqnarray}
with $u=\frac{F}{B}$. After some manipulations we find an upper
bound on the Hilbert-Schmidt norm of $G'(t)$
$$
\|G'(t)\|_{HS}\leq \frac{C\, e^{\delta\, t}}{|\sin^3(Bt)|}
$$
where $\gamma<\delta<0$ and the constant $C$ is uniform in $t$ and
depends on $f,g,F,B$. The last inequality yields the following
estimate
$$
 \left \| \sum_{n\in\N_0}\int_{n\pi/B+\eps}^{(n+1)\pi/B-\eps}G(t)e^{-i\lambda
t} \D t \right \|
 \leq |\lambda|^{-1}\left [2\, \|f\|_{\infty}\, \|g\|_{\infty} +
C(\delta)\int_{\eps}^{\pi/B-\eps}\frac{1}{|\sin^3(Bt)|}
 \, \D t\right ]
$$
here we have put
$$
C(\delta) = \frac{C\, e^{\delta\eps/2}}{1-e^{\delta\pi/2B}}\; , \qquad (\delta<0)
$$
Finally, we can sum up all the contributions on the r.h.s. of
(\ref{resolventa}) to write
\begin{eqnarray} \label{finalest}
\|f(H_1(F)-\lambda-i\, \gamma)^{-1}g\| & \leq & \|f\|_{\infty}\,
\|g\|_{\infty}\left\{ \left
(1+\frac{2e^{-\gamma\epsilon}}{1-e^{\gamma\pi/B}}
\right )\, \eps +2|\lambda|^{-1}\right \} \nonumber \\
& + &
C(\delta)|\lambda|^{-1}\int_{\eps}^{\pi/B-\eps}\frac{1}{|\sin^3(Bt)|}\,
\D t
\end{eqnarray}
Sending $\eps$ to zero in a suitable way, for example as
$|\lambda|^{-\alpha}$ with $\alpha>0$ and sufficiently small, we
can make sure that the last term in (\ref{finalest}) tends to zero
as $\lambda\ra\pm\infty$ and the claim of the
 Proposition then follows. The case $\gamma>0$ can be proved in a
 similar way.
\end{proof}

Armed with Proposition \ref{limit} we can prove the promised
result about the finite number of eigenvalues in the vicinity of
real axis.

\begin{prop} \label{finite}
Suppose that assumptions $(b)$ and $(c)$ hold true. Then for any
$aF<bF<\beta_0$ there exists some $M(a)$ such that $H(F,ib)$ has
no eigenvalues in the strip $S_a:=\{0\geq \Im z\geq -aF,\, |\Re
z|\geq M(a)\}$.
\end{prop}

\begin{proof}
We write $V_1:=|V(x+ib,y)|^{1/2}, \quad V_2:=|V(x+ib,y)|^{1/2}\,
{\rm phase}V(x+ib,y)$ and, for $z\in S_a$,
$R_1(z)=(z-H_1(F,ib))^{-1}, \quad R(z)=(z-H(F,ib))^{-1}$. Then, by
an approximation argument and Proposition \ref{limit}
\begin{equation} \label{limzero}
\lim_{\lambda\ra\pm\infty}\|V_1(H_1(F,ib)-\lambda-i\gamma)^{-1}V_2\|=0,
\quad \gamma>F(b-a)>0
\end{equation}
so that the Neumann series
$$
R(z) = \sum_{n=0}^{\infty}R_1(z)(VR_1(z))^n=R_1(z)+R_1(z)V_1\left
(\sum_{n=0}^{\infty} (V_2R_1(z)V_1)^n\right )V_2 R_1(z)
$$
converges for $z\in S_a$. Moreover, since $\|R_1(z)\|\leq
((b-a)F)^{-1}$, we can conclude that
$$
\sup_{z\in S_a}\|(z-H(F,ib))^{-1}\|<\infty
$$
\end{proof}

The following definition is a ``translational version'' of the
notion of analytic vectors for dilatation group introduced in
\cite{AC}.

\begin{defin}
Let $A$ be any open complex domain having non-empty intersection
with $\R$. Then we denote by $\mathcal{D}(A)$  a set of those
vectors $f$, for which $f_{\theta}=U(\theta)f,\, \theta\in\R$ can
be analytically continued to $A$.
\end{defin}

We are now able to state the main theorem of this section. Since a
similar analysis was made in \cite{Herbst} for a non magnetic
case, we skip some details of the proof referring to the latter.

\begin{thm} \label{lifetime}
Take $\alpha:=\alpha_0 F>0$ sufficiently small such that the conditions
$(a),\, (b)$ and $(c)$ are satisfied for $\min
(\beta,\beta_0)>\alpha$. Assume moreover that $bF>\alpha$ and let
$\psi,\, \phi,\, H_1(F)\psi,\, H_1(F)\phi \in\mathcal{D}(\{z\in
\C:\, |\Im z|\leq bF\})$. Then for any $t\geq 0$

$$
(\psi,e^{-i\, tH(F)}\phi)=\sum_{-\Im E_j\leq\alpha}
(\psi_{-ib},P_j(ib)\phi_{ib})\, e^{-i\, tE_j}+R(t)
$$
where
$$
R(t)\leq {\rm {\cal C}}\, e^{-t(\alpha+\eps)}
$$ for some
$\eps>0$. Here $P_j(ib)$ is the spectral projector of $H(F,ib)$
associated with the eigenvalue $E_j$.
\end{thm}

\begin{proof}
Following \cite{Herbst} we put $K_1(z)=(\psi,(z-H(F))^{-1}\phi)$
for $\Im z>0$ and note that $K_1(z)$ has a meromorphic
continuation to $\C$, which is for $\Im z>-bF$ given by
$K_1(z)=(\psi_{-ib},(z-H(F,ib))^{-1}\phi_{ib})$. Similarly
$K_2(z)=(\psi,(z-H(F))^{-1}\phi)$, $\Im z<0$ has for $\Im z<bF$ a
meromorphic continuation given by
$K_2(z)=(\psi_{ib},(z-H(F,-ib))^{-1}\phi_{-ib})$. \par From the
spectral theorem it follows that
\begin{equation} \label{int}
(\psi,e^{-i\, tH(F)}\phi)=\int_{-\infty}^{\infty}Q(\lambda)\, e^{-
it\lambda}\D\lambda
\end{equation}
where $Q(\lambda)$ is the spectral density. We have
\begin{eqnarray}
Q(\lambda) & = & \lim_{\delta\ra 0}\,
\frac{i}{2\pi}(\psi,[\lambda+i\delta-H(F))^{-1}-(\lambda-i\delta-H(F))^{-
1}]\phi)
\nonumber \\
& = &-(2\pi i)^{-1}(K_1(\lambda)-K_2(\lambda)), \quad \lambda\in\R
\end{eqnarray}
Let us now take $a$ such that $\alpha<aF<bF$. By Proposition
\ref{finite} and assumption $(c)$, the meromorphic continuation of
$Q(\lambda)$ to $\C$, which is given by
$$
Q(z)= -(2\pi i)^{-1}(K_1(z)-K_2(z))
$$
is then analytic in the strip $S_a$ and on the real axis. In
addition, the argument of \cite{Herbst} shows that for
$0<\gamma<aF$ and $|E|$ large enough
\begin{equation} \label{l1}
Q(E-i\gamma)=\mathcal{O}(|E|^{-2})
\end{equation}
This allows us to shift the integration in (\ref{int}) from the
real axis downwards to the lower complex half-plane by
$$
\lambda\ra\lambda-i\, (\alpha+\eps) \qquad \alpha+\eps<aF
$$
so that
\begin{eqnarray}
(\psi,e^{-i\, tH(F)}\phi)& = & 2\pi i\sum_{-\Im E_j\leq\alpha}\,
{\rm Res}\,
K_1(z)|_{z=E_j}\, e^{-itE_j} \nonumber \\
& + &
e^{-t(\alpha+\eps)}\int_{-\infty}^{\infty}Q(\lambda-i(\alpha+\eps))\,
e^{-it\lambda}\D\lambda
\end{eqnarray}
For the residues of $K_1(z)$ we have
$$
{\rm Res}\, K_1(z)|_{z=E_j}=\frac{1}{2\pi i}\int_{|z-
E_j|=\varepsilon} \D z(\psi_{-ib},(z-H(F,ib))^{-1}\phi_{ib})
=(\psi_{-ib},P_j(ib)\phi_{ib})
$$
However, $f_j(z)=(\psi_{\bar{z}},P_j(z)\phi_{z})$ is by assumption
an analytic function of $z$ for $-F\Im z<\Im E_j$. Since $f_j(z)$
is constant for $z$ real, we can conclude that $f_j(z)$ is
independent of $z$ as long as $-F\Im z<\Im E_j$.
\end{proof}

\section{Green function of $H_1(F,ib)$} \label{GreenFunction}

\noindent As already announced, we now proceed to the estimations
of the Green function of the crossed fields Hamiltonian $H_1(F,ib)$.
Results of this Section have a technical character and will be
used in the announced forthcoming paper, in which we prove an
upper bound on the resonance widths.

\subsection{General solution}

 \par We want to find an upper bound on
the Green function (and its first derivatives) of \be H_1(ib):= H_1(F,ib) =
-\pd_x^2+(-i\pd_y-Bx)^2 -Fx-Fib \ee Since $H_1(ib)$ is
translationally invariant in $y-$direction, it can be written as
\be H_1(ib) \simeq \int_{\R}^{\oplus}H_1(ib,k)\D k \ee where \be H_1(ib,k)=
-\pd_x^2+(k-Bx)^2 -Fx-Fib \ee is the corresponding fiber
Hamiltonian on $L^2(\R,\D x)$. Its spectral equation \be
H_1(ib,k)\psi(x,k)=z \psi(x,k) \ee can be solved explicitly to
give two linearly independent solutions. Namely, with the notation
\be
 x(k):= x-\frac{k}{B}\,-\frac{F}{2B^2}, \quad z(k):= z+ibF+\frac{F}{B}\,
 k+\,\frac{F^2}{4B^2}
\ee we get for $x(k)>0$: \bea \psi_1(x,k) &=& e^{-Bx^2(k)/2}\,
U\left( \frac{B-z(k)}{4B}, \frac{1}{2},
  B\, x^2(k)\right ) \\
\psi_2(x,k) &=& e^{-Bx^2(k)/2}\, V\left( \frac{B-z(k)}{4B},
\frac{1}{2},
  B\, x^2(k)\right )  \\
&=& e^{-Bx^2(k)/2}\sqrt{\pi}\,
\left[\frac{M\left(\frac{B-z(k)}{4B},
      \frac{1}{2},B\, x^2(k)\right)}{\Gamma\left(\frac{3B-z(k)}{4B}\right
)}+
      2\sqrt{B}\, x(k)\, \frac{M\left(\frac{3B-z(k)}{4B},
      \frac{3}{2},B\, x^2(k)\right)}{\Gamma\left(\frac{B-z(k)}{4B}\right
    )}\right ] \nonumber
\eea and for  $x(k)\leq0$: \bea \psi_1(x,k) &=& e^{-Bx^2(k)/2}\,
V\left( \frac{B-z(k)}{4B}, \frac{1}{2},
  B\, x^2(k)\right ) \\
\psi_2(x,k) &=& e^{-Bx^2(k)/2}\, U\left( \frac{B-z(k)}{4B},
\frac{1}{2},
  B\, x^2(k)\right )
\eea where $U$ and $M$ are solutions to Kummer's equation,  see
\cite[chap. 13]{AS}. Here we have followed the analysis made in
\cite{ejk} for purely magnetic Hamiltonian. Clearly,
\mbox{$V\left( (B-z(k))/4B, 1/2, B\, x^2(k)\right)$} is analytical
continuation of $U\left( (B-z(k))/4B, 1/2, B\, x^2(k)\right)$ for
$x(k)<0$. We note that $\psi_1(x,k)\in L^2([0,\infty))$ and
$\psi_2(x,k)\in L^2((-\infty,0])$. The Green function of
$H_1(ib,k)$ is thus given by \be G(x,x';z,k)=
\frac{\psi_1(x_>,k)\, \psi_2(x_<,k)}{W(\psi_1,\psi_2)} \ee with
\be x_>=\max(x,x'), \quad x_<= \min(x,x') \ee With the help of
\cite[p. 505]{AS} one can calculate the Wronskian \be
W(\psi_1,\psi_2)=\sqrt{\pi B}\, 2^{\frac{3}{2}-\frac{z(k)}{2B}}\,
\Gamma^{-1}\left( \frac{B-z(k)}{2B}\right ) \ee
\par
Green's function of $H_1(ib)$ then reads \be \label{greenfunc}
G_1({\bf x},{\bf x}';z)= (\pi
B)^{-1/2}\int_{\R}2^{-\frac{3}{2}+\frac{z(k)}{2B}}\psi_1(x_>,k)\,
\psi_2(x_<,k)\, \Gamma\left( \frac{B-z(k)}{2B}\right )\,
e^{ik(y-y')}\, \D k \ee

To discuss the convergence of the integral in the definition of
$G_1({\bf x},{\bf x}';z)$ we recall the behaviour of the
hypergeometric functions $U$ and $M$, see \cite[p. 504]{AS}. The
latter gives the asymptotic of the integrand in (\ref{greenfunc})
in the form:
$$
e^{-k[|x'-x|-i(y'-y)]}\, \left (\frac{x-kB^{-1}}{x'-kB^{-1}}\right
  )^{\frac{z(k)}{2B}}\,
  \frac{1}{\sqrt{(x-kB^{-1})(x'-kB^{-1})}}\,\,
 [1+\mathcal{O}(k^{-2})]
$$
as $k\ra\infty$, and
$$
e^{k[|x'-x|-i(y'-y)]}\, \left (\frac{x'-kB^{-1}}{x-kB^{-1}}\right
  )^{\frac{z(k)}{2B}}\,
  \frac{1}{\sqrt{(x-kB^{-1})(x'-kB^{-1})}}\,\,
 [1+\mathcal{O}(k^{-2})]
$$
as $k\ra-\infty$. Thus, for $x'\neq x$ the integral converges
independently on the value of $y',y$, for in that case the
asymptotic is given by \be\label{kinf} e^{-|k||x'-x|}\,
\alpha(k)^k\, k^{-1}, \quad |k|\ra\infty \ee with
$\lim_{|k|\to+\infty}\alpha(k)=1$. Similarly, when $y'\neq y$ the
integral converges even for $x'=x$, since the asymptotic then
reads \be \label{kinf2} e^{-ik(y'-y)}\,
  \frac{1}{\sqrt{(x-kB^{-1})(x-kB^{-1})}}\,\,
 [1+\mathcal{O}(k^{-2})],\quad |k|\ra\infty,
\ee and simple integration by parts shows that $G_1({\bf x},{\bf
x}';z)$ converges pointwise for any $y'\neq y$.
\\
\noindent From the definition of hypergeometric functions and the
construction of $\psi_1$ and $\psi_2$ it follows, that the product
$\psi_1(x,k)\, \psi_2(x,k)$ is analytic w.r.t. $k$. The integrand
of (\ref{greenfunc}) is thus a meromorphic function with poles at
\be k_2=-BF^{-1}(z_2+bF),\quad
k_1(n)=BF^{-1}\left[(2n+1)B-z_1-F^2/(4B)\right], n\geq 0 \ee where
we write $k=k_1+ik_2$ and $z=z_1+iz_2$. Moreover the integrand
vanishes in the limit $|k_1|\ra\infty$, see (\ref{kinf}),
(\ref{kinf2}). Therefore we can shift the integration to the lower
complex half-plane by substituting \be \label{subs} p:=
-\frac{k}{B}\, -\frac{F}{2B^2}\, -i\, \frac{z_2+bF}{2F} \,  \delta
\quad,\quad \delta=\frac{y-y'}{|y-y'|}, \ee so that \be x(p) = x+p+i\,
\Delta, \quad x'(p) = x'+p+i\, \Delta, \quad \Delta =
\frac{z_2+bF}{2F}\, \delta \ee

Since $U(a,b,t)$ is a many-valued function with a principal branch
$-\pi<\arg t\leq \pi$,  we have to consider its analytical
continuation, see \cite[p. 504]{AS}. The fundamental solutions
$\psi_1(x_>,p)$ and $\psi_2(x_<,p)$ will be given by different
combinations of hypergeometric functions corresponding to
different values of quasimomentum $p$;
\begin{enumerate}
\item For $p<-x'<-x$:
\bea \psi_1(x',p) &=& e^{-Bx'^2(p)/2}\, V\left( \frac{B-z(p)}{4B},
\frac{1}{2},
  B\, x'^2(p)\right ) \\
\psi_2(x,p) &=& e^{-Bx^2(p)/2}\, U\left( \frac{B-z(p)}{4B},
\frac{1}{2},
  B\, x^2(p)\right )
\eea \item For $-x'<p<-x$: \bea \psi_1(x',p) &=& e^{-Bx'^2(p)/2}\,
U\left( \frac{B-z(p)}{4B}, \frac{1}{2},
  B\, x'^2(p)\right ) \\
\psi_2(x,p) &=& e^{-Bx^2(p)/2}\, U\left( \frac{B-z(p)}{4B},
\frac{1}{2},
  B\, x^2(p)\right )
\eea \item For $-x'<-x<p$: \bea \psi_1(x',p) &=& e^{-Bx'^2(p)/2}\,
U\left( \frac{B-z(p)}{4B}, \frac{1}{2},
  B\, x'^2(p)\right ) \\
\psi_2(x,p) &=& e^{-Bx^2(p)/2}\, V\left( \frac{B-z(p)}{4B},
\frac{1}{2},
  B\, x^2(p)\right )
\eea
\end{enumerate}

The Cauchy theorem now yields \bea \label{greenf} && G_1({\bf
x},{\bf x}';z)= (\pi B)^{-1/2} e^{-\frac{z_2+bF}{2F}\,
  |y-y'|}\, e^{-iF(y-y')/2+B(z_2+bF)^2/(4F)}  \\
&&\times
\int_{\R}2^{-\frac{3}{2}+\frac{z(k(p))}{2B}}\psi_1(x',k(p))\,
\psi_2(x,k(p)) \, \Gamma\left( \frac{B-z(k(p))}{2B}\right )\,
e^{ipB(y'-y)} \, \D p \nonumber \eea
with $k(p)$ defined through (\ref{subs}). \\

\subsection{Long distances: $G_1({\bf x},{\bf x}';z)$}

Let us suppose, for definiteness, that $x'>x$ and examine the case
where $|x'-x|>1$. For $x$ and $x'$ we have to consider the
following three cases: $x'>x>0$, $x'>0>x$ and $0>x'>x$. In each
case we perform the integral \eqref{greenf} by dividing it in
several pieces depending on the value of $p$. Before doing so we
give some general estimates on the hypergeometric functions which
will be used throughout the text. \\

\begin{rem} The symbol $C$ below denotes a
positive real number, which depends on the energy $z$, but not on
the size of the electric field $F$. \end{rem}

\noindent For the product $U(a,b,t)\, M(a,b,t)$ we use the
asymptotic expressions, \cite[p. 504]{AS}, and the corresponding
estimate of the error term to get \bea\label{estim1} && \left
|2^{-\frac{3}{2}+\frac{z(p)}{2B}} \, V\left( \frac{B-z(p)}{4B},
\frac{1}{2},
  B\, x'^2(p)\right ) \, U\left( \frac{B-z(p)}{4B}, \frac{1}{2},
  B\, x^2(p)\right )\Gamma\left( \frac{B-z(p)}{2B}\right ) \right | \leq
 \nonumber \\
&& C\, e^{B x'^2(p)} \left |\frac{p+x+i\Delta}{p+x'+i\Delta}\right
|^{z(p)/2B}\, B^{-1/2}|(x+p+i\Delta)(x'+p+i\Delta)|^{-1/2}\,
[1+C\Delta^{-2}] \eea where we have used the doubling formula for
the gamma function, \cite[p. 256]{AS} \be \Gamma(2z)=
\pi^{-\frac{1}{2}}\, 2^{2z-1}\, \Gamma(z)\, \Gamma(z+\tfrac{1}{2})
\ee \noindent Henceforth we will work only with the leading term
and drop the factor $[1+C\Delta^{-2}]$. Moreover, as the
asymptotic behaviour of both summands in the definition of $V$ is
identical, we will consider only the first one. \par The following
bound can be easily found \be
|(x+p+i\Delta)(x'+p+i\Delta)|^{-1/2}\leq \Delta^{-1}\; . \ee We
have
 \bea
 \left|\frac{p+x+i\Delta}{p+x'+i\Delta}\right|^{z(p)/2B}&=&
\left(1+\frac{(x-x')^2}{(p+x')^2+\Delta^2}+
\frac{2(x'-x)(p+x')}{(p+x')^2+\Delta^2}\right)^{\frac{\tilde{z}_1-Fp}{4B}}
\eea with $\tilde{z}_1=z_1-F^2/4B^2$. Remark that $|\cdots|>1$,
thus for $\tilde{z}_1\leq 0$ and $p\geq 0$ this term can be
neglected. For $\tilde{z}_1>0$ we can apply the following
inequality \be 1+\frac{(x-x')^2}{(p+x')^2+\Delta^2}+
\frac{2(x'-x)(p+x')}{(p+x')^2+\Delta^2} \leq
1+\frac{2(x-x')^2}{\Delta^2}\; . \ee For $p<0$ we write
$|\cdots|^{-\frac{Fp}{2B}}=e^{-\frac{Fp}{2B}\, \ln|\cdots|}$.
Finally, note that the same result holds true if we interchange
$x$ and $x'$, which correspond to interchange the functions $U$
and
$V$.\\

\emph{\underline{Let $x'>x>0$}}\\

\noindent We divide the interval of integration in five parts as
follows
$$\mathbb{R}=(-\infty,-2x']\cup (-2x',-x']\cup (-x',-x]\cup (-x,-x/2]\cup
(-x/2,\infty)$$

\noindent {\it For $p\in (-\infty,-2x']$}:\\
Keeping in mind that $F\to 0$ one gets from \eqref{estim1} \bea &&
\int_{-\infty}^{-2x'}\left |\, 2^{-\frac{3}{2}+\frac{z(p)}{2B}}\,
\psi_1(x',x,p)\, \psi_2(x',x,p) \, \Gamma\left(
\frac{B-z(p)}{2B}\right ) \right | \D p
\nonumber \\
&\leq& \frac{C}{\sqrt{B}}\, \Delta^{-1} \left [1+\,
\frac{2(x'-x)^2}{\Delta^2}\right]^{\frac{z_1}{4B}}
e^{\frac{B}{2}\, (x'^2-x^2)}\int_{-\infty}^{-2x'} e^{p\,
B(x'-x)}\, e^{\frac{-Fp}{4B}\, \ln|\cdots|} \D p
\nonumber \\
&\leq& \frac{C}{\sqrt{B}}\, \Delta^{-1} \left[1+\,
\frac{2(x'-x)^2}{\Delta^2}\right]^{\frac{z_1}{4B}}\,
e^{\frac{B}{2}\, (x'^2-x^2)}\int_{-\infty}^{-2x'} e^{p
B(x'-x)/2}\, \D p
\nonumber \\
&\leq& \frac{C}{B^{3/2}}\Delta^{-1}\, \left
  [1+\, \frac{2(x'-x)^2}{\Delta^2}\right]^{\frac{z_1}{4B}}
  e^{-\, \frac{B}{2}\,(x'-x)^2}\,
\eea

\noindent {\it For $p\in (-x/2,\infty)$}:\\
\eqref{estim1} (with $x$ and $x'$ interchanged) and the bounds
given before lead to \bea\label{firstbound} &&
\int_{-x/2}^{\infty}\left |2^{-\frac{3}{2}+\frac{z(p)}{2B}}\,
\psi_1(x',x,p)\, \psi_2(x',x,p) \Gamma\left(
\frac{B-z(p)}{2B}\right ) \right | \D p
\nonumber \\
&\leq& \frac{C}{\sqrt{B}}\Delta^{-1}\left
  [1+\, \frac{2(x'-x)^2}{\Delta^2}\right]^{\frac{z_1}{4B}}
 e^{\frac{B}{2}(x^2-x'^2)} \times \nonumber \\
 &\times&\left\{\int_{-x/2}^{0} e^{-Bp(x'-x)}e^{-\frac{Fp}{2B}\ln |\ldots|}
\D p  + \int_{0}^{\infty} e^{-Bp(x'-x)}\D p \right\}
 \nonumber \\
&\leq& \frac{C}{\sqrt{B}}\Delta^{-1}\left
  [1+\, \frac{2(x'-x)^2}{\Delta^2}\right]^{\frac{z_1}{4B}}
 e^{\frac{B}{2}(x^2-x'^2)}
 \left\{\int_{-x/2}^{0} e^{-2Bp(x'-x)}
\D p + \int_{0}^{\infty} e^{-Bp(x'-x)} \D p \right\}\nonumber \\
&\leq& \frac{C}{B^{3/2}}\Delta^{-1}\left
  [1+\, \frac{2(x'-x)^2}{\Delta^2}\right]^{\frac{z_1}{4B}}
 2e^{-\,\frac{B}{2}\, (x'-x)^2}
 \eea

\noindent {\it For $p\in (-2x',-x']$}:\\
Here the estimate (\ref{estim1}) does not give us the sought
result. Instead we will rewrite the corresponding part of the
integration in (\ref{greenf}) in the following way, \bea
\label{phi} && \int_{-2x'}^{-x'} \left |\,
2^{-\frac{3}{2}+\frac{z(p)}{2B}}\, \psi_1(x',x,p)\, \psi_2(x',x,p)
\, \Gamma\left( \frac{B-z(p)}{2B}\right ) \right | \D p \nonumber \\
&& \equiv  \Delta^{-1}\, x'^{-1}\, (x'-x)^{\frac{z_1}{2B}} e^{-\,
\frac{B}{4}\, (x'-x)^2}\int_{-2x'}^{-x'} \Phi(x',x,p) \D p \eea
and look at the maximum of the function $\Phi(x',x,p)$ in the
interval $[-2x',-x']$. We denote the maximum value by
$\Phi_0(x',x)$. In particular we want to show that $\Phi_0$ is
bounded above by certain function of $F$, which does not grow
faster than a power function of $F^{-1}$ as $F\ra 0$. To be more precise, we want to show,
that there exist some positive constants $\Theta_0,\, \theta_1$, such that
$$
|\Phi(x',x,p)| \leq \Theta_0\, F^{-\theta_1}
$$
holds uniformly for $p\in (-2x',-x']$ and $F$ small enough. This procedure
will used below also for other values of $p$. \par
We recall the asymptotic properties of the gamma function, see
\cite[p. 257]{AS} \be \label{asympgamma} \Gamma(az+b) \sim
\sqrt{2\pi}\, e^{-az}\, (az)^{az+b-\frac{1}{2}},\quad
|z|\ra\infty,\,\, |\arg z|<\pi,\,\, a>0 \ee It is then easy to
see, that $\Phi(x',x,p)$ is bounded at the endpoints of the
interval $[-2x',-x']$. We can thus confine ourselves to the case
when $\Phi$ acquires its maximum inside the considered interval.
Let us denote the corresponding extremal point by
$$
p_0(x') = -x'-j(x')
$$
First of all we note that if $j(x')$ is bounded, one can show the
boundedness of $\Phi(x',x,p_0(x'))$ in the same way as that of
$\Phi(x',x,-x')$. Without loss we may thus assume that $j(x')$ is
unbounded. We shall distinguish two different situations according
to different behaviour of the function $j(x')$.
\begin{enumerate}
\item $j^2(x')/x'$ bounded as
$x'\ra\infty$. In this case the first parameter of
\be\label{largeM} M\left(\frac{B-z(p_0(x'))}{4B}, \frac{1}{2}, B\,
x'^2(p_0(x'))\right ) \ee does not grow more slowly than its
argument, for \bea z(p_0(x'))&=&z_1+F(x'+j(x'))-\frac{F^2}{4B^2}\,
+\frac{i}{2}\,
(z_2+bF)(2-\delta) \\
B\, x'^2(p_0(x')) &=& B\, (j(x')+i\Delta)^2. \eea We observe that
in our case real parts of $z(p_0(x'))$ and $x'^2(p_0(x'))$
increase faster than their imaginary parts in the limit
$x'\ra\infty$. It then follows from the definition of function
$M$, \cite[p. 504]{AS}, that the behaviour of (\ref{largeM}) at
infinity will be governed by \be M\left(\frac{B-\Re
z(p_0(x'))}{4B}, \frac{1}{2}, \Re B\, x'^2(p_0(x'))\right ) \ee
The application of a suitable asymptotic expansion, \cite[p.
105]{buch}, also \cite[p. 509, 13.5.21]{AS}, thus gives us the
following inequality for $x'\ra\infty$ \be \label{Mexp} \left|\,
M\left(\frac{B-\Re z(p_0(x'))}{4B}, \frac{1}{2},  \Re B\,
    x'^2(p_0(x'))\right )\right|
\leq C\, F^{-1}\, e^{\frac{j^2(x')}{2}} \ee Recalling
(\ref{asympgamma}) we can conclude that \bea && \Phi(x',x,p_0(x'))
\leq C\, \Delta\, x'\, \exp{\left[ -\frac{B}{4}\,
  \left((x'-x)^2+2j^2(x')+4j(x')(x'-x)\right)\right]} \nonumber \\
&& |\, B(x'-x+j(x'))|^{\frac{F(x'+j(x'))}{2B}}\, \left |\,
\Gamma\left(
    \frac{B-z(p_0(x'))}{4B}\right)\right |
\eea is bounded above by a constant times $\Delta\, F^{-1}$.

\item $j^2(x')/x'$ unbounded. Here we can use
again
  (\ref{estim1}) and the boundedness of $\Phi(x',x,p_0(x'))$ then follows
  after some elementary manipulations.
\end{enumerate}

\noindent To sum up we have \bea \label{2x'} && \int_{-2x'}^{-x'}
\left |\, 2^{-\frac{3}{2}+\frac{z(p)}{2B}}\, \psi_1(x',x,p)\,
\psi_2(x',x,p)
\, \Gamma\left( \frac{B-z(p)}{2B}\right ) \right | \D p \nonumber \\
&& \leq C\,  (F^{-1}+\Delta^{-1}) \, (x'-x)^{\frac{z_1}{2B}}
e^{-\, \frac{B}{4}\, (x'-x)^2} \eea

\noindent {\it For $p\in (-x,-x/2]$}:\\
Same estimations as for $p\in (-2x',-x']$.\\

\noindent {\it For $p\in (-x',-x]$}:\\
We show that the function to be integrated is bounded by some
constant uniform in $x,x'$ times $e^{-\frac{B}{4}(x-x')^2}$. At
the boundary it has been shown above that the function is bounded,
we suppose that there is an extremal point
$p_0=p_0(x,x')\in(-x',x]$. Denote
$$
d(x,x')=|p_0+x| \qquad \text{ and } \qquad d'(x,x')=|p_0+x'|
$$
the distances between the end points and the extremum $p_0$.\\
We have to consider the following cases, which correspond to the
different behaviours of the argument of $U$: $d(x,x')$ unbounded,
$d(x,x')< C$ and
the same for $d'(x,x')$. \\

\noindent 1) $d(x,x'),\, d'(x,x')$ unbounded: we have for $p=p_0$
\bea\label{xx'>0} {\cal A}_1(x,x')& := &
e^{\frac{B}{4}(x+p_0+i\Delta)^2}\,
\sqrt{|W^{-1}\, (\psi_1,\psi_2)|}\, |\psi_1(x,p)| \\
&=&\left|2^{\frac{z(p_0)}{4B}}e^{-\frac{B}{4}(x+p_0+i\Delta)^2}
B(x+p_0+i\Delta)^\frac{z(p)-B}{2B}\right|\left|\Gamma\left(\frac{B-
z(p_0)}{2B}\right)\right|^{1/2}
 \nonumber\\
&\leq&2^{\frac{\tilde{z}_1-Fp_0}{4B}}e^{-\frac{B}{4}(x+p_0)^2}
(B\left|x+p_0+i\Delta\right|)^\frac{\tilde{z}_1-
B}{2B}\left|\Gamma\left(\frac{B-\tilde{z}_1-Fp_0}{2B} +
i\eta\right)\right|^{1/2} \nonumber
\end{eqnarray}
where $\eta$ denote the imaginary part of the argument in the
gamma function. ${\cal A}_2(x,x')$ is defined in the same way
where $\psi_1$ is replaced with $\psi_2$ and $x,x'$ are
interchanged.
In the limit $x',x\ra\infty$ we consider the following cases.\\
$a)$ \be \label{larged} B(d^2(x,x')+\Delta^2),\,\,
B(d'^2(x,x')+\Delta^2) > \nu_0 \frac{\tilde{z}_1-Fp_0}{4B}: \ee
where $\nu_0=4(1+\ln 2)f_0^{-1}>1$ and $f_0>0$ is the global
minimum of $(1-t\ln (2/t))$ for $t\geq 0$. Using the asymptotic
properties of the gamma function we get for the leading term of
\eqref{xx'>0}: \be \label{leadingterm}
\exp\left\{-\frac{B}{4}(x+p_0)^2\left[1+f(x,x') \ln
\left(-2f^{-1}(x,x')\right)\right]+(1+\ln 2)\,
\frac{\tilde{z}_1-Fp_0}{4B} \right\}
\end{equation}
where \be f(x,x')=\frac{F\, p_0(x,x')}{B^2\, (x+p_0(x,x'))^2}<0
\ee The boundedness of ${\cal A}_1(x,x')$ follows from
(\ref{larged}). Same analysis for ${\cal A}_2(x,x')$ then gives
\bea \label{1a} && \left |\, \psi_1(x',x,p)\, \psi_2(x',x,p) \,
W^{-1}(\psi_1,\psi_2) \right | \leq
e^{-\frac{B}{4}(x+p_0)^2}e^{-\frac{B}{4}(x'+p_0)^2} {\cal A}_1
{\cal A}_2
\nonumber \\
&& \leq C\, e^{-\, \frac{B}{8}\, (x'-x)^2} \eea

To continue we recall again the asymptotic behaviour of
$U(a,b,z)$, see \cite[p. 504]{AS}, to assure that \bea
\label{estU} && \left |\, U\left(\frac{B-z(p_0(x',x))}{4B},
\frac{1}{2}, B\,
    (x'+p+i\Delta)^2)\right )\right| \leq \nonumber \\
&& C\, \left |\,  U\left(\frac{B-\Re z(p_0(x',x))}{4B},
\frac{1}{2},  B\,
    ((x'+p)^2+\Delta^2)\right )\right| \, [1+C\Delta^{-2}]
\eea Let us now consider
\\
$b)$
$$
B(d^2(x,x')+\Delta^2) > \nu_0\, \frac{\tilde{z}_1-Fp_0}{4B},\quad
B(d'^2(x,x')+\Delta^2)= \nu\, \frac{\tilde{z}_1-Fp_0}{4B},\, \,
\nu\in[1,\nu_0],
$$
in which case the part corresponding to ${\cal A}_1(x,x')$ can be
treated as above and for the rest of the integrand we use \cite[p.
509, 13.5.20]{AS} to get \be \left |e^{-\frac{B}{2}(x'+p_0)^2}
U\left(\frac{B-\Re z(p_0(x'))}{4B},
    \frac{1}{2},  B\, ((x'+p)^2+\Delta^2)\right ) \right | \leq C\, e^{-
\frac{B}{4\nu}\, (x'+p_0)^2} \ee and consequently \bea && \left
|\, \psi_1(x',x,p)\, \psi_2(x',x,p) \, W^{-1}(\psi_1,\psi_2)
\right | \leq
e^{-\frac{B}{4}(x+p_0)^2}e^{-\frac{B}{4\nu}(x'+p_0)^2} {\cal A}_1
\nonumber \\
&& \leq C\, e^{-\, \frac{B}{8\nu}\, (x'-x)^2} \eea $c)$ \be
\label{1c} B(d^2(x,x')+\Delta^2)\geq
\frac{\tilde{z}_1-Fp_0}{4B},\quad B(d'^2(x,x')+\Delta^2)<
\frac{\tilde{z}_1-Fp_0}{4B}, \ee The part which includes
$\psi_1(x,p)$ can be controlled by one of the estimates given
above. For the second part we observe that, \cite[p. 509,
13.5.22]{AS}, $|\psi_2(x',p)|$ is uniformly bounded for $p$ in
$(-x',-x]$. The properties of gamma function then lead to the
following inequality for the Wronskian \bea \label{wronskian}
|W^{-1/2}\, (\psi_1,\psi_2)|&& \leq C\, \exp \left
[\frac{Fp_0}{4B}(\ln(\sqrt{-Fp_0/2B})-1-\ln 2)\right ]\,
e^{\frac{Fp_0}{4B}\, \ln(\sqrt{-Fp_0/2B})}\, \left
|\frac{Fp_0}{2B}\right|^{-\frac{\tilde{z}_1}{4B}}   \nonumber \\
&& \leq C\, \exp
[-B((x'+p_0)^2+\Delta^2)(\ln(\sqrt{(x'+p_0)^2+\Delta^2})-1-\ln2)]
\nonumber
\\
&& \leq C\, e^{-B(x'+p_0)^2}, \eea so that \bea && \left |\,
\psi_1(x',x,p)\, \psi_2(x',x,p) \, W^{-1}(\psi_1,\psi_2) \right |
\leq e^{-\frac{B}{4\nu}(x+p_0)^2}\, |W^{-1/2}(\psi_1,\psi_2)|
\nonumber \\
&& \leq C\, e^{-\, \frac{B}{8\nu}\, (x'-x)^2} \eea $d)$
$$
B(d^2(x,x')+\Delta^2)< \frac{\tilde{z}_1-Fp_0}{4B},\quad
B(d'^2(x,x')+\Delta^2)< \frac{\tilde{z}_1-Fp_0}{4B}
$$
Here both the functions $|\psi_2(x',p)|$ and $|\psi_1(x,p)|$ are
uniformly bounded and the exponential decay then comes from the
Wronskian in the
same way as in the case $c)$. \\

\noindent 2) One of $d(x,x'),\, d'(x,x')$ bounded.\par Let us
suppose for definiteness, that $d(x,x')$ is bounded. At the point
$p=p_0(x,x')$ we apply again (\ref{estU}) and \cite[p. 508,
13.5.16]{AS} to find that \be \label{estU2}
 |\psi_1(x,p)|\leq C\, \left |\Gamma \left( \frac{1}{2} -
   \frac{B-z(p_0(x',x))}{4B}\right )\right |
\ee For the function $\psi_2(x',p)$ and for the Wronskian we use
the suitable estimate given above in one of the cases $a),\, b),\,
c),\, d)$, which gives the desired result. \par In all these cases
the same analysis can be made when $d(x,x')$ and $d'(x,x')$
interchange their roles. \\

\noindent 3) Both $d(x,x')$ and $d'(x,x')$ bounded.\par Since this
can only happen when $|x'-x|\leq C$ , it suffices to show that the
integrand is bounded. The latter however follows immediately from
(\ref{estU2}) and
$$
\left |\Gamma^2 \left( \frac{1}{2} -
   \frac{B-z(p_0(x',x))}{4B}\right )\, W^{-1}(\psi_1,\psi_2) \right |\leq
C,\quad \forall\, p\in (-x',-x]
$$

\noindent Finally we conclude that there exists certain constant
$\omega>0$, which depends on $B$ but not on $F$, such that \bea &&
\int_{-x'}^{-x} \left |\, \psi_1(x',x,p)\, \psi_2(x',x,p) \,
W^{-1}(\psi_1,\psi_2) \right | \D p \leq
\int_{-x'}^{-x}e^{-\frac{B}{4}(x+p_0)^2}e^{-\frac{B}{4}(x'+p_0)^2}
{\cal A}_1
  {\cal A}_2 \D p
\nonumber \\
&& \leq C\, \Delta^{-1}\, (x'-x)\, e^{-\omega\, (x'-x)^2} \eea

\begin{rem} 
We do not present the analysis of all the possible combinations,
because the in the remaining cases one can proceed in a completely
analogous way as above.
\end{rem}

\emph{\underline{Let $x'>0>x$}}\\

\noindent In this case we divide the interval of integration in
four parts as
$$\mathbb{R}=(-\infty,-2x']\cup (-2x',-x']\cup (-x',-x]\cup (-x,\infty)$$

\noindent The intervals $(-\infty,-2x']$, $(-2x',-x']$ can be
treated exactly as in the previous case. For $p\in (-x,\infty)$ we
proceed in the same way as for $p\in (-x/2,\infty)$ in the
previous case, keeping in mind that since $x<0$ one has $p>0$.\\

\noindent \emph{For $p\in (-x',-x]$} we separate the analysis of
the integrand
in two pieces.\\
\noindent {(1) $p\in (-x',0]$}: Same argument as for the interval
$(-x',-x]$ when $x',x$ are both positive.

\noindent {(2) $p\in (0,-x]$}: We divide the interval in
$(0,p_c+1]\cup(p_c+1,-x]$, where $p_c=\frac{\tilde{z}_1-B}{F}$.
For $p>p_c$ we have $\Re a(p)>0$ with $a(p)$ the first parameter
of the function $U$. In this case we can use the integral
representation of $U$ to get \cite{DMP4} \be\label{intrepr}
|U(a(p),\tfrac{1}{2},\rho(p))|\leq \frac{C}{\Re
a(p)}|\Gamma(a(p))|^{-1} \qquad \text{for } \;\Re \rho(p)>0,
\;\;\Re a(p)>0
 \ee
In $(0,p_c+1]$ the analysis of the maximum of
$$
|x'+p+i\Delta|^2|x+p+i\Delta|^2
$$
shows that it is a power function in $(x'-x)$. Thus, since the
$\Gamma$ function remains in this interval bounded, we get the
bound
$e^{-\frac{B}{2}(x'-x)^2}$ times a polynomial in $(x'-x)$.\\
In $(p_c+1,-x-|\Delta|]$ we use the bounds \eqref{intrepr} and the
asymptotic behaviour of the gamma function to get a uniform upper
bound. In $(-x-|\Delta|,-x]$ we use \eqref{intrepr} for the
function $U$ depending on $x'$ while for the other $U$ we use its
expression in term of a sum of function $M$. In this case we get
a uniform estimate since the argument of $M$ is bounded. \\

\emph{\underline{Let $0>x'>x$}}\\

\noindent We divide the interval of integration in four parts as
follows
$$\mathbb{R}=(-\infty,0]\cup (0,-x']\cup (-x',-x]\cup (-x,\infty)$$

\noindent For the interval $(-x,\infty)$ the remarks above hold.
When $p\in (-x',-x]$ a slight modification of the analysis done in
$(0,-x]$ above leads to the desired bound.

\noindent {\it For $p\in (-\infty,0]$}: \bea &&
\int_{-\infty}^{0}\left |\, 2^{-\frac{3}{2}+\frac{z(p)}{2B}}\,
\psi_1(x',x,p)\, \psi_2(x',x,p) \, \Gamma\left(
\frac{B-z(p)}{2B}\right ) \right | \D p
\nonumber \\
&\leq& \frac{C}{\sqrt{B}}\, \Delta^{-1} \left [1+\,
\frac{2(x'-x)^2}{\Delta^2}\right]^{\frac{z_1}{4B}}
e^{\frac{B}{2}\, (x'^2-x^2)}\int_{-\infty}^{0} e^{p\, B(x'-x)}\,
e^{\frac{-Fp}{4B}\, \ln|\cdots|} \D p
\nonumber \\
&\leq& \frac{C}{\sqrt{B}}\, \Delta^{-1} \left[1+\,
\frac{2(x'-x)^2}{\Delta^2}\right]^{\frac{z_1}{4B}}\,
e^{\frac{B}{2}\, (x'^2-x^2)}\int_{-\infty}^{0} e^{p B(x'-x)/2}\,
\D p
\nonumber \\
&\leq& \frac{C}{B^{3/2}}\Delta^{-1}\, \left
  [1+\, \frac{2(x'-x)^2}{\Delta^2}\right]^{\frac{z_1}{4B}}
  e^{-\, \frac{B}{2}\,(x'-x)^2}\, \eea

\noindent {\it For $p\in (0,-x']$}:
 \bea && \int_{0}^{-x'}\left
|\, 2^{-\frac{3}{2}+\frac{z(p)}{2B}}\, \psi_1(x',x,p)\,
\psi_2(x',x,p) \, \Gamma\left( \frac{B-z(p)}{2B}\right ) \right |
\D p
\nonumber \\
&\leq& \frac{C}{\sqrt{B}}\, \Delta^{-1} \left [1+\,
\frac{2(x'-x)^2}{\Delta^2}\right]^{\frac{z_1}{4B}}
e^{\frac{B}{2}\, (x'^2-x^2)}\int_{0}^{-x'} e^{p\, B(x'-x)}\,
 \D p
\nonumber \\
&\leq& 2\frac{C}{B^{3/2}}\Delta^{-1}\, \left
  [1+\, \frac{2(x'-x)^2}{\Delta^2}\right]^{\frac{z_1}{4B}}
  e^{-\, \frac{B}{2}\,(x'-x)^2}\, \eea

\vspace{0.5cm} \noindent Let us finally formulate the results in
\begin{lem}\label{greenestimate}
For $F$ small enough and $|x'-x|\geq 1$ there exist some strictly
positive
 constants $C_1,\, C_2,\, \tilde{\omega}$, which depends on $B$ and $z$,
such that the following inequality holds true \be |G_1({\bf
x},{\bf x};z)|\leq C_1\, \Delta^{-1} e^{-\Delta\,
  |y-y'|}\, e^{-\tilde{\omega}\, (x'-x)^2}\, \left
[1+\, \frac{2(x'-x)^2}{\Delta^2}\right]^{\frac{z_1}{4B}}\,
[1+C_2\Delta^{- 2}] \ee with $\Delta=\frac{z_2+bF}{2F}$.
\end{lem}

\subsection{Long distances: $\pd_{x,y}\, G_1({\bf x},{\bf x}';z)$}

In this section we want to prove similar result to that one
described in Lemma \ref{greenestimate} also for the derivatives of
the Green function w.r.t. $x$ and $y$. We suppose again that
$x'>x$ and $|x'-x|>1$. As we have already seen the most general
and complicated case is the one where $x',x>0$ and the all the
others can be regarded as its simplification. Therefore here we confine
ourselves to the situation when both $x',x$ are positive. \par We
start with the derivative w.r.t. $x$. For $|x'-x|>1$ the integral
$$
\int_{\R} \left |\, 2^{-
\frac{3}{2}+\frac{z(p)}{2B}}\psi_1(x',p)\, \pd_x \psi_2(x,p) \,
\Gamma\left( \frac{B-z(p)}{2B}\right )\right |\, \D p
$$
converges uniformly with respect to $x$, see (\ref{kinf}). We can
thus interchange the differentiation and integration in (\ref
{greenf}) to get the following inequality for the derivative of
$G_1({\bf x},{\bf x};z)$: \bea \label{diffgreen}
&& |\pd_x\, G_1({\bf x},{\bf x};z)|\leq \\
&& C\, e^{-\Delta |y'-y|}\, \int_{\R} \left |\, 2^{-
\frac{3}{2}+\frac{z(p)}{2B}}\psi_1(x',p)\, \pd_x \psi_2(x,p) \,
\Gamma\left( \frac{B-z(p)}{2B}\right )\right |\, \D p  \nonumber
\eea We split again the integration in (\ref{greenf}) into five
intervals:
$$
\mathbb{R}=(-\infty,-2x']\cup (-2x',-x']\cup (-x',-x]\cup
(-x,-x/2]\cup (-x/2,\infty)
$$
and use \cite[p. 507, 13.4.8/21]{AS} to calculate the derivatives
of hypergeometric functions. When $p\in(-x/2,\infty)$ we get for
the corresponding integrand in (\ref{diffgreen}) \bea &&
-B(x+p+i\Delta)\, \psi_1(x',p)\, \psi_2(x,p)\,
W^{-1}(\psi_1,\psi_2)
+ 2B(x+p+i\Delta) e^{-B(x+p+i\Delta)^2/2}\, a(p)\sqrt{\pi} \nonumber \\
&& \times\Bigg [ \frac{M(a(p)+1,
      \frac{3}{2},B(x+p+i\Delta)^2)}{\frac{1}{2}\, \Gamma(a(p)+1/2)}
   +2\sqrt{B}(x+p+i\Delta)\,
\frac{M(a(p)+\frac{3}{2},\frac{5}{2},B(x+p+i\Delta)^2)}{\frac{3}{2}\,\Gamma
(a(p))}
\nonumber \\
&& +2\sqrt{B}\, \frac{M(a(p)+1,\frac{3}{2},
    B(x+p+i\Delta)^2)}{a(p)\Gamma(a(p))} \Bigg ] \psi_1(x',p)\, W^{-
1}(\psi_1,\psi_2) \eea where \be a(p) = \frac{B-z(p)}{4B} \; . \ee
The first term can be controlled in the same way as the Green
function itself due to (\ref{estim1}) and the fact that
\be\label{basic} \left |\frac{x+p+i\Delta}{x'+p+i\Delta} \right
|^2 \leq 1+\frac{2(x-x')^2}{\Delta^2} \ee As for the term which
includes the derivative of the function $M$, using \cite[p.
504]{AS} and $\Gamma(a+1)=a\Gamma(a)$, we note that the asymptotic
behaviour of \be \frac{a(p)\, M(a(p)+1,\frac{3}{2},
B(x+p+i\Delta)^2)}{\Gamma(a(p)+1/2)}\, W^{-1}(\psi_1,\psi_2) \ee
is the same as that of \be \frac{M(a(p),\frac{1}{2},
B(x+p+i\Delta)^2)}{\Gamma(a(p)+1/2)}\, W^{-1}(\psi_1,\psi_2) \ee
The rest of the analysis is then identical with the case of
$G_1({\bf x},{\bf x}';z)$ itself.
\par
For $p<-x'$ are $x,x'$ interchanged and we have to differentiate
the function $U$: \be \label{derivU} \pd_x
U\left(a(p),\frac{1}{2}, B(x+p+i\Delta)^2\right )=-2B
(x+p+i\Delta)\, a(p)\, U\left (a(p)+1,\frac{3}{2},
B(x+p+i\Delta)^2 \right ) \ee The pre-factor $(x+p+i\Delta)$ is
again well controlled due to (\ref{basic}). In addition we observe
that for the product \be a(p)\, U\left (a(p)+1,\frac{3}{2},
B(x+p+i\Delta)^2\right )\, V\left
  (a(p),\frac{1}{2}, B(x+p+i\Delta)^2\right )
\ee we get the upper bound (\ref{estim1}) multiplied by \be
\label{factor}
 \left |\frac {a(p)}{(x+p+i\Delta)^2} \right |
\ee and that for $p<-2x'$ is the latter uniformly bounded w.r.t.
to $x,x'$. Thus, for $x\in (-\infty,-2x']$ we can use the same
estimations as for $G_1({\bf x},{\bf x}';z)$.
\par
For $p\in (-2x',-x'] \cup (-x,-x/2]$ we multiply the function
$\Phi (x',x,p)$ introduced in (\ref{phi}) by $a(p)$, which leads
to an additional factor $F^{-1}$ in the estimate (\ref{2x'}). \par

Similarly is for $p\in (-x',-x]$ the factor (\ref{factor}), coming
from the derivative of $U$, controlled by the decay of the upper
bounds that we have found above. More exactly, for the case $1a)$
we see from the inequality (\ref{larged}) that (\ref{factor}) is
uniformly bounded in the interval $(-x',-x]$. The case $1b)$ is
treated in an analogous way. As for $1c)$, we note that
$$
a(p_0)\, e^{-\frac{B}{8}\, (x+p_0)^2}
$$
is bounded due to (\ref{1c}). The result then follows from
(\ref{1a}). When the inequalities of the case $1d)$ hold, then
following (\ref{wronskian}) we get
$$
| W^{-1}(\psi_1,\psi_2)\, a(p_0)|\leq C e^{-\frac{B}{2}\,
(x'+p_0)^2}\, e^{-\frac{B}{2}\, (x+p_0)^2},
$$
which gives again the exponential decay of the integrand. In the
cases $2)$ and $3)$ we proceed in the same way as for the Green
function itself noting that both
$$
\left |a(p_0) \Gamma(1/2-a(p_0)) W^{-1/2}(\psi_1,\psi_2)\right
|,\quad \left
  |a(p_0) \Gamma^2(1/2-a(p_0)) W^{-1}(\psi_1,\psi_2)\right |
$$
are uniformly bounded. We thus conclude that \be \left |\pd_x\,
\psi_1(x',x,p)\, \psi_2(x',x,p) \, W^{-1}(\psi_1,\psi_2) \right |
\leq C\, e^{-\, \frac{B}{16\nu}\, (x'-x)^2} \ee for $p\in
(-x',-x]$.\par Same arguments can be then used for $\pd_y\,
G_1({\bf x},{\bf x}';z)$. Since the substitution $k\ra p$ is not
analytic in $y$, the differentiation w.r.t. $y$ has to be done
before this substitution is made. In other words, we have to
differentiate the formula (\ref{greenfunc}) and then substitute
$p$ for $k$ through (\ref{subs}). This leads to a multiplication
of the integrand in (\ref{diffgreen}) by the factor $Bp$, which is
well controlled by the previously given arguments, noting that
$$
\left |\frac {p}{\sqrt{(x+p+i\Delta)(x'+p+i\Delta)}} \right |
$$
is uniformly bounded on $(-\infty,-2x']\cup (-x/2,\infty)$. \par
Finally we get
\begin{lem}\label{diffgreenestimate}
For $F$ small enough and $|x'-x|\geq 1$ there exist some strictly
positive
 constants $C_3,\, C_4,\, \tilde{\omega}$, which depends on $B$ and $z$,
such that the following inequality holds true \be |\pd_{x,y}\,
G_1({\bf x},{\bf x}';z)|\leq C_3\, F^{-2}\, \Delta^{-1} e^{-
\Delta\,
  |y-y'|}\, e^{-\tilde{\omega}\, (x'-x)^2}\, \left
[1+\,
\frac{2(x'-x)^2}{\Delta^2}\right]^{\frac{z_1}{4B}+\frac{1}{4}}\,
[1+C_4\Delta^{-2}] \ee with $\Delta=\frac{z_2+bF}{2F}$.
 \end{lem}

\subsection{Short distances}

Up to now we have considered that $|x'-x|\geq 1$ and  $|y'-y|$ was
arbitrary. Here we want to investigate the case where $|x'-x|< 1$
for any value of $|y'-y|$. Since our system is two-dimensional, we
expect the Green function $G_1({\bf x},{\bf x}';z)$ to have a
logarithmic singularity as $x\ra x'$ and $y\ra y'$ of the
following type:
$$
G_1({\bf x},{\bf x}';z) \sim \ln (|{\bf x}'-{\bf x}|)
$$
Our goal in this section is to show that \be
\int_{\R}\int_{|x'-x|\leq 1} |\pd_{x,y}^n G_1({\bf x},{\bf x}';z)|
e^{\frac{\Delta}{2}|y-y'|} \D x' \D y' \quad n=0,1 \ee is bounded
as a function of $x$ and $y$. We will work only with the
derivatives of $G_1({\bf x},{\bf x}';z)$, noting that same
arguments then apply also to $G_1({\bf x},{\bf x}';z)$ itself.
\par We divide the real
axis as above and present again only the case $x',x>0$. \\

\emph{\underline{$\pd_x G_1({\bf x},{\bf x}';z)$} }\\

\noindent From the asymptotic expansion for the integrand of
$G_1({\bf x},{\bf x};z)$, see (\ref{kinf}), (\ref{kinf2}), it follows that
$$
\int_{\R} |\pd_x\, \psi_1(x',x,p)\, \psi_2(x',x,p)\,
W^{-1}(\psi_1,\psi_2) |\D p
$$
converges only if $x'\neq x$. This reflects the usual behaviour of
the Green function, i.e. the discontinuity of the derivative for
$x'=x$. We will thus investigate $\pd_x G_1({\bf x},{\bf x}';z)$
separately for $(x'-x)$ in the compacts of $(0,1)$ and $(-1,0)$.
\par

Assume first that $(x'-x)\in (0,1)$. For the derivative w.r.t. $x$
we write \be |\pd_x\, G_1({\bf x},{\bf x}';z)| = C\,
e^{-\Delta|y'-y|}\, \left |\int_{\R}
  g(x',x,p)\, e^{ipB(y'-y)} \D p \right |
\ee where for $p>-x$ \be g(x',x,p) = \psi_1(x',p)\, \pd_x
\psi_2(x,p)\, W^{-1}(\psi_1,\psi_2) \ee Let us perform first the
integration in the interval $p\in(-x/2,\infty)$. We have \bea &&
\pd_x \psi_2(x,p) =-B(x+p+i\Delta)\, \psi_2(x,p) +
e^{-\frac{B}{2}\,
  (x+p+i\Delta)^2}\,\pd_x V\left(a(p),\frac{1}{2},B (x+p+i\Delta)^2\right )
\nonumber \\
&& =: \phi_1(x,p) + \phi_2(x,p) \eea Using the asymptotic
expansions for $M$ and $U$ and integrating by parts we find \bea
\label{firstint} && \left |\int_{-x/2}^{\infty} \psi_1(x',p)\,
\phi_1(x,p)\, W^{-1}(\psi_1,\psi_2)\, e^{ipB(y'-y)} \D p \right |
= C\, e^{-B(x'^2-
x^2)/2} \\
&& \times \Bigg |\int_{-x/2}^{\infty}
  e^{-pB[(x'-x)-i(y'-y)]}\, \left (\frac{p+x'+i\Delta}{p+x+i\Delta}\right
  )^{\frac{z(p)}{2B}}\,
  \frac{p+x+i\Delta}{\sqrt{(p+x+i\Delta)(p+x'+i\Delta)}} \nonumber \\
&&
[1+\mathcal{O}(|p+x+i\Delta|^{-2})][1+\mathcal{O}(|p+x'+i\Delta|^{-2})]]
 \D p \Bigg | \nonumber \\
&& \leq \frac{C}{B|(x'-x)-i(y'-y)|}\, \left [\Delta^{-1}+
e^{-B(x'^2- x^2)/2}\, \int_{-x/2}^{\infty}
  e^{-pB(x'-x)}\, w(x',x,p) \D p \right ]\, [1+C\, \Delta^{-2}]  \nonumber
\eea where \be w(x',x,p) = \pd_p \left \{ \left
(\frac{p+x'+i\Delta}{p+x+i\Delta}\right
  )^{\frac{z(p)}{2B}}\,
  \frac{p+x+i\Delta}{\sqrt{(p+x+i\Delta)(p+x'+i\Delta)}}\right \}
\ee Here we have used the fact that the integrand of
(\ref{firstint}) is an analytic function of $p$ and therefore we
can differentiate the term
$$
[1+\mathcal{O}(|p+x+i\Delta|^{-2})][1+\mathcal{O}(|p+x'+i\Delta|^{-2})]]
$$
w.r.t. $p$. It then follows from the Cauchy formula, that the
derivative is an $L^1[(-x/2,\infty)]$ function with the
corresponding norm smaller than a constant times $\Delta^{-1}$.
The first term on the last line of (\ref{firstint}) gives the
expected result. The point is now that, as one can easily verify,
the function $w(x',x,p)$ is proportional to $(x'-x)$ in the sense
that
$$
\frac{w(x',x,p)}{x'-x}
$$
is uniformly bounded. In other words \be \left |
e^{-B(x'^2-x^2)/2}\, \int_{-x/2}^{\infty} e^{-pB(x'-x)}\,
w(x',x,p) \D p \right | \leq C \ee and \be \left
|\int_{-x/2}^{\infty} \psi_1(x',p)\, \phi_1(x,p)\,
W^{-1}(\psi_1,\psi_2)\, e^{ipB(y'-y)} \D p \right |\leq \frac{C\,
\Delta^{-1}}{|(x'-x)-i(y'-y)|}\, [1+C\, \Delta^{-2}] \ee All
constants in the latter inequality are uniform for $(x'-x)$ in the
compacts of $(0,1)$. \par \noindent Same analysis can be made also
for the term $\phi_2(x,p)$, which includes the derivative of the
function $M$, see the remarks below (\ref{basic}). \par For $p$ in
the interval $(-\infty,-2x']$ are $x'$ and $x$ interchanged and we
have \be g(x',x,p) = \psi_2(x',p)\, \pd_x \psi_1(x,p)\,
W^{-1}(\psi_1,\psi_2) \ee so that $\phi_1(x,p)$ is unchanged and
instead of $\phi_2(x,p)$ we get \be \tilde{\phi}_2(x,p)=
e^{\frac{B}{2}\,
  (x+p+i\Delta)^2}\,\pd_x U\left(a(p),\frac{1}{2},B (x+p+i\Delta)^2\right )
\ee Using (\ref{derivU}) and (\ref{factor}) we can proceed as
above replacing $w(x',x,p)$ with \bea
&& \tilde{w}(x',x,p) = w(x',x,p)\, \frac {a(p)}{(x+p+i\Delta)^2} \\
&&\, + \left (\pd_p\, \, \frac {a(p)}{(x+p+i\Delta)^2}\right )\,
\left
  (\frac{p+x+i\Delta}{p+x'+i\Delta}\right)^{\frac{z(p)}{2B}}\,
  \frac{p+x+i\Delta}{\sqrt{(p+x+i\Delta)(p+x'+i\Delta)}} \nonumber
\eea It is now sufficient to realize that \be \pd_p \left (\frac
{a(p)}{(x+p+i\Delta)^2}\right )\in L^1((-\infty,-2x']) \ee with
the corresponding $L^1$ norm being uniformly bounded from above by
a constant times $\Delta^{-1}$, and that \be e^{\frac{pB}{2}\,
(x'-x)}\, \left (\frac{p+x+i\Delta}{p+x'+i\Delta}\right
  )^{\frac{z(p)}{2B}}\,
  \frac{p+x+i\Delta}{\sqrt{(p+x+i\Delta)(p+x'+i\Delta)}}
\ee is uniformly bounded for $p\in (-\infty,-2x']$ provided $F$ is
small enough. This follows from \be \ln\left |\,
\frac{p+x+i\Delta}{p+x'+i\Delta} \right | \leq C, \quad \forall\,
p\in(-\infty,-2x'] \ee Then \bea \label{secondint} && \left
|\int_{-\infty}^{-2x'} \psi_1(x,p)\, \tilde{\phi}_2(x',p)\,
W^{-1}(\psi_1,\psi_2)\, e^{ipB(y'-y)} \D p \right | \\
&& \leq \frac{C}{B|(x'-x)-i(y'-y)|}\, \left [\Delta^{-1}+
e^{B(x'^2- x^2)/2}\, \int_{-\infty}^{-2x'}
  e^{pB(x'-x)}\, \tilde{w}(x',x,p) \D p \right ]\, [1+C\, \Delta^{-2}]
\nonumber \\
&& \leq \frac{C\, \Delta^{-1}}{|(x'-x)-i(y'-y)|}\, [1+C\,
\Delta^{-2}] \nonumber \eea uniformly for $(x'-x)$ in the compacts
of $(0,1)$, since both \be e^{B(x'^2-x^2)/2}\,
\int_{-\infty}^{-2x'} e^{pB(x'-x)}|w(x',x,p)| \D p,\quad
e^{B(x'^2-x^2)/2} e^{-Bx'(x'-x)} \ee
are bounded. Same bounds on $|\pd_x G_1({\bf x},{\bf x};z)|$ can be found for $(x'-x)\in (-1,0)$. \\

\emph{\underline{$\pd_y G_1({\bf x},{\bf x}';z)$ }}\\

\noindent As it was already noticed, differentiation w.r.t. $y$
leads to a multiplication of the corresponding integrand by the
factor $iBp$: \be |\pd_y\, G_1({\bf x},{\bf x}';z)| = C\,
e^{-\Delta|y'-y|}\, \left |\int_{\R}
  h(x',x,p)\, e^{ipB(y'-y)} \D p \right |
\ee where for $p>-x$ \be h(x',x,p) = iBp\, \psi_1(x',p)\,
\psi_2(x,p)\, W^{-1}(\psi_1,\psi_2) \ee and for $p<-x'$ \be
h(x',x,p) = iBp\, \psi_1(x,p)\, \psi_2(x',p)\,
W^{-1}(\psi_1,\psi_2). \ee We can thus proceed in the same way as
for $\pd_x\, G_1({\bf x},{\bf
  x}';z)$. The only new ingredient which we need is the fact that
that \be \left (\pd_p\, \,
\frac{p}{\sqrt{(p+x+i\Delta)(p+x'+i\Delta)}} \right)\in L^1\left
((-\infty,-2x']\cup (-x/2,\infty)\right ), \ee where the $L^1$
norm is again bounded by a constant times $\Delta^{-1}$. \par For
$p\in(-2x',-x/2]$ we apply to both $\pd_x G_1({\bf x},{\bf x}';z)$
and $\pd_y G_1({\bf x},{\bf x}';z)$  the same arguments as for
$|x'-x|\geq 1$ noting that these are independent on the value of
$(x'-x)$.

\noindent We have thus proved

\vspace{0.3cm}

\begin{lem}\label{shortdist}
For $F$ small enough there exists some strictly positive constant
$G'_0$ such that the following inequality holds true \be
\int_{\R}\int_{|x'-x|< 1} |\pd_{x,y}^m G_1({\bf x},{\bf
x}';z)|e^{\frac{\Delta}{2}|y-y'|} \, {\rm d} x' {\rm d} y' \leq
G'_0\, \Delta^{-3}, \ee where $m=0,1$.
\end{lem}

\appendix

\section{Integral kernel of $e^{-i tH_1}$ }

Here we sketch the calculation of the integral kernel of evolution
operator $e^{-i\, tH_1}$ in the gauge $H_L=p_x^2+(p_y-Bx)^2$. We
employ the functional integration to write
\begin{equation} \label{kernel}
(x,y|e^{-i\, tH_1}|x_0,y_0) =
\int_{x_0,y_0;0}^{x,y;t}\D[w(\cdot)]\exp\left \{i\, \int_0^t\D\,
s\, L[w(s),\dot{w}(s)]\right \}
\end{equation}
where
$$
  L[w(s),\dot{w}(s)] =\frac{1}{4}\, |\dot{w}(s)|^2 +Fw_x(s)-
\dot{w}_y(s)Bw_x(s)
$$
is the Lagrangian and
\begin{equation}
S_t[w(\cdot)] = \int_0^t\D\, s\, L[w(s),\dot{w}(s)]
\end{equation}
the corresponding action. The integral in (\ref{kernel}) is then
taken over all trajectories $w(s)$ which satisfy the boundary
conditions
\begin{equation} \label{bc}
w(0)=(x_0,y_0), \quad w(t)=(x,y)
\end{equation}
We will write $w$ as a sum of a classical trajectory plus certain
fluctuation:
$$
w(s) = w_{cl}(s) + \xi(s)
$$
and evaluate $S_t[w(\cdot)]$ in the vicinity of the classical
action $S_t[w_{cl}(\cdot)]$. As $L[w(s),\dot{w}(s)]$ is a
quadratic function of canonical variables, all higher variations
of $S_t[w_{cl}(\cdot)]$ are identically zero and
\begin{equation} \label{action}
S_t[w(\cdot)] = S_t[w_{cl}(\cdot)] + \delta^{(1)}
S_t[w_{cl}(\cdot)] + \delta^{(2)} S_t[w_{cl}(\cdot)]
\end{equation}
Moreover, since $w_{cl}(s)$ minimises the classical action, the
second term on the r.h.s. of (\ref{action}) vanishes and for the
last term we have
$$
\delta^{(2)} S_t[w_{cl}(\cdot)] = \int_0^t \D\,
s\left\{\frac{1}{4}\,
|\dot{\xi}(s)|^2-\dot{\xi_y}(s)B\xi_x(s)\right \}
$$
From the Van Vleck formula it then follows that the kernel
(\ref{kernel}) can be expressed in terms of the classical action
only:
\begin{equation} \label{kernel2}
(x,y|e^{-i\, tH_1}|x_0,y_0) = \frac{1}{2\pi i}\, e^{i\,
S_t[w_{cl}(\cdot)]}\left [\det\left \{-\frac{\pd^2
S_t[w_{cl}(\cdot)]}{\pd \alpha \pd \beta_0}\right
\}_{\alpha,\beta}\right ]^{1/2}
\end{equation}
with $\alpha,\beta \in \{x,y\}$.\\
To compute $S_t[w_{cl}(\cdot)]$ we have to find the solution of
the classical equations of motion
\begin{eqnarray} \label{class}
\frac{1}{2}\, \ddot{w}_x^{cl} & = & -B\dot{w}_y^{cl}+F\nonumber \\
\frac{1}{2}\, \ddot{w}_y^{cl} & = & B\dot{w}_x^{cl}
\end{eqnarray}
It is not difficult to verify that the general solution of
(\ref{class}) reads
\begin{eqnarray}
w_x^{cl}(s) & = & C_1(t)\cos(2Bs)+C_2(t)\sin(2Bs)+C_3(t) \nonumber \\
w_y^{cl}(s) & = & -C_2(t)\cos(2Bs)+C_1(t)\sin(2Bs)+u\, s+B^{-1}\,
C_4(t)
\end{eqnarray}
where $u =\frac{F}{B}$ is the drift velocity in $y-$direction and
the ``constants'' $\{C_i(t),i=1,2,3,4\}$ depend on $t$ through the
boundary conditions (\ref{bc}). A straightforward calculation
gives
\begin{eqnarray}
w_x^{cl}(s) & = & \tfrac{1}{2}\, \left
[(y-y_0-ut)+(x-x_0)\cot(Bt)\right
]\, \sin(2Bs) \nonumber \\
& - & \tfrac{1}{2}\, \left [(x-x_0) - (y-y_0-ut)\cot(Bt)\right ]\,
\cos(2Bs) \nonumber \\
& + & \tfrac{1}{2}\, \left [(x+x_0) - (y-y_0-ut)\cot(Bt)\right ]
\end{eqnarray}
and similarly
\begin{eqnarray}
w_y^{cl}(s) & = & -\tfrac{1}{2}\, \left
[(x-x_0)-(y-y_0-ut)\cot(Bt)\right
]\, \sin(2Bs) \nonumber \\
& - & \tfrac{1}{2}\, \left [(y-y_0-ut)+(x-x_0)\cot(Bt)\right ]\,
\cos(2Bs)
\nonumber \\
& + & \tfrac{1}{2}\, \left [(y+y_0-ut)+(x-x_0)\cot(Bt)\right ]
+u\, s
\end{eqnarray}
The action then takes the form
\begin{eqnarray}
S_t[w_{cl}(\cdot)] & = & \tfrac{1}{4}\tfrac{F^2}{B^2}\, t
+\tfrac{1}{2}\tfrac{F}{B}\,
\left(y-y_0-\tfrac{F}{B}t\right)-\tfrac{1}{2}\,
B(x+x_0)\left(y-y_0-\tfrac{F}{B}t\right) \nonumber \\
& + & \tfrac{1}{4}\, B\cot(Bt)\left
[\left(y-y_0-\tfrac{F}{B}t\right)^2+(x- x_0)^2\right ]
\end{eqnarray}
and Van Vleck's determinant is thus easily calculated to give the
integral kernel of $e^{-i\, tH_1}$
\begin{equation} \label{evol}
(x,y|e^{-i\, tH_1}|x_0,y_0) = \frac{1}{2\pi i}\,
\sqrt{\tfrac{B}{2}}\, e^{i\, S_t[w_{cl}(\cdot)]}\,
\frac{1}{\sin(Bt)}
\end{equation}

\section*{Acknowledgements}
We wish to thank P.A.Martin and N.Macris for suggesting to us the
presented problem and for many stimulating and encouraging
discussions throughout the project. Numerous comments of P.Exner
are also gratefully acknowledged. H.K. would like to thank his
hosts at Institute for Theoretical Physics, EPF Lausanne for a
warm hospitality extended to him. C.F. thanks the Math. department
of Stuttgart University, where the part of the present work was
done for hospitality. The work of C.F. was supported by the Fonds
National Suisse de la Recherche Scientifique No. 20-55694.98.

\end{document}